\documentclass{article}
\usepackage{psfig}
\usepackage{latexsym}
\usepackage{amssymb}
\usepackage{epsfig}
\usepackage[USenglish]{babel}

 \newtheorem{lemma}{Lemma}
 \newtheorem{theorem}{Theorem}
 \newcommand{\proof}{\noindent\bf Proof: \rm}
 \newcommand{\qed}{$\Box$}
\newcommand{\old}[1]{{}}
\newcommand{\eps}{\varepsilon}
\newcommand{\ol}[1]{\overline{#1}}
\newcommand{\OPT}{\mbox{\it OPT}}
\newcommand{\HEU}{\mbox{\it HEU}}
\newcommand{\OPTij}{\mbox{\it OPT}_{ij}}
\newcommand{\OPTip}{\mbox{\it OPT}_{i\bullet}}
\newcommand{\OPTop}{\mbox{\it OPT}_{0\bullet}}
\newcommand{\OPTpj}{\mbox{\it OPT}_{\bullet j}}

\newcommand{\notOPTip}{\ol{\mbox{\it OPT}}_{i\bullet}}
\newcommand{\notOPTop}{\ol{\mbox{\it OPT}}_{0\bullet}}
\newcommand{\notOPTpj}{\ol{\mbox{\it OPT}}_{\bullet j}}
\newcommand{\HEUij}{\mbox{\it HEU}_{ij}}
\newcommand{\HEUip}{\mbox{\it HEU}_{i\bullet}}
\newcommand{\HEUpj}{\mbox{\it HEU}_{\bullet j}}

\newcommand{\notHEUip}{\ol{\mbox{\it HEU}}_{i\bullet}}
\newcommand{\notHEUpj}{\ol{\mbox{\it HEU}}_{\bullet j}}

\newcommand{\bi}{\begin{itemize}}
\newcommand{\ei}{\end  {itemize}}
\newcommand{\bt}{\begin{tabbing}}
\newcommand{\et}{\end  {tabbing}}
\newcommand{\be}{\begin{enumerate}}
\newcommand{\ee}{\end  {enumerate}}

\title{Maximum Dispersion\\ and Geometric Maximum Weight Cliques\thanks{
A preliminary version of this paper appears in the proceedings
of APPROX 2000, pp.\ 132--143.}}
\author{
{S\'andor P.\ Fekete}\thanks{Department of Optimization,
Braunschweig University of Technology,
38106 Braunschweig,
GERMANY,
{\tt fekete@tu-bs.de}}
\and
{Henk Meijer}\thanks{
Department of Computing and Information Science, 
Queen's University, 
Kingston, Ontario K7L 3N6, CANADA,
{\tt henk@cs.queensu.ca}.
Research partially supported by NSERC.}
}

\date{}

\begin{document}
\maketitle

\begin{abstract}
We consider a
facility location problem, where the objective is to 
``disperse'' a number of facilities, i.e., select
a given number $k$ of locations from a discrete set of $n$ candidates,
such that the average distance between selected locations
is maximized. 
In particular, we present algorithmic 
results for the case where vertices are 
represented by points in $d$-dimensional space, 
and edge weights correspond to rectilinear 
distances.  Problems of this type have been considered 
before, with the best result being an 
approximation algorithm with performance ratio 2. 
For the case where $k$ is fixed, we establish a 
linear-time algorithm that finds an optimal 
solution. For the case where $k$ is part of 
the input, we present a polynomial-time 
approximation scheme.
\end{abstract}

\section{Introduction}
A common problem in the area of facility location is the
selection of a given number of $k$ locations from a set $P$
of $n$ feasible positions, such that the selected set
has optimal distance properties. Natural objective functions
are the maximization of the minimum distance, or 
of the average distance between selected
points; {\em dispersion} problems of this type come into play whenever
we want to minimize interference between the corresponding facilities.
Examples include oil storage tanks, ammunition dumps, nuclear power 
plants, hazardous waste sites, and fast-food outlets 
(see \cite{rrt-hscadp-94,chandra01approximation}). In the latter paper,
the problem of maximizing the average distance
is called the {\em Remote Clique} problem.

Formally, problems of this type can be described as follows:
given a graph $G=(V,E)$ with $n$ vertices, and non-negative
edge weights $w_{v_1,v_2}=d(v_1,v_2)$ for $(v_1,v_2)\in E$. 
Given $k\in \{2,\ldots,n\}$,
find a subset $S\subset V$ with $|S|=k$, such that 
$w(S):=\sum_{(v_i,v_j)\in E(S)} d(v_i,v_j)$ is maximized.
(Here, $E(S)$ denotes the edge set of the subgraph of $G$ 
induced by the vertex
set $S$.) 

{From} a graph theoretic point of view, this problem has
been called a {\em heaviest subgraph problem}. Being a
weighted version of a generalization of the problem
of deciding the existence of a $k$-{\em clique}, i.e.,
a complete subgraph with $k$ vertices, the problem is
strongly NP-hard \cite {t-oflg-91}.
It should be noted that H{\aa}stad \cite{h-chawn-99} showed that 
the problem {\sc Clique} of maximizing the cardinality of
a set of vertices with a maximum possible number of edges 
is in general hard to approximate within $n^{1-\varepsilon}$.
For the heaviest subgraph problem, we want to maximize the number
of edges for a set of vertices of given cardinality,
so H{\aa}stad's result does not imply an immediate performance bound.

\subsection*{Related Work}
Over recent years, there have been a number of approximation
algorithms for various subproblems of this type. Feige and 
Seltser \cite{fs-dksp-97} have studied the graph problem 
(i.e., edge weights are $0$ or $1$) and showed how to find
in time $n^{O((1+\log \frac{n}{k})/\varepsilon)}$
a $k$-set $S\subset V$ with $w(S)\geq 
(1-\varepsilon)\left(\begin{array}{c}k\\2\end{array}\right)$,
provided that a $k$-clique exists. They also gave evidence that
for $k\simeq n^{1/3}$, semidefinite programming fails to 
distinguish between graphs that have a $k$-clique, and
graphs with densest $k$-subgraphs having average degree
less than $\log n$.

Kortsarz and Peleg
\cite{kp-cds-93} describe a polynomial algorithm with
performance guarantee $O(n^{0.3885})$ for the general case where
edge weights do not have to obey the triangle inequality.
A newer algorithm by Feige, Kortsarz, and Peleg~\cite{feige01dense}
gives an approximation ratio of $O(n^{1/3}\log n)$.
For the case where $k=\Omega(n)$, 
Asahiro, Iwama, Tamaki, and Tokuyama
\cite{aitt-gfdg-96} give a greedy constant factor approximation,
while Srivastav and Wolf \cite{sw-fdssp-98} use semidefinite
programming for improved performance bounds. For the case
of dense graphs (i.e., $|E|=\Omega(n^2)$) and $k=\Omega(n)$,
Arora, Karger, and Karpinski
\cite{akk-ptasdinhp-95} give a polynomial time approximation scheme.
On the other hand, Asahiro, Hassin, and Iwama~\cite{AHI}
show that deciding the existence of a ``slightly dense''
subgraph, i.e., an induced subgraph on $k$ vertices that has at least 
$\Omega(k^{1+\varepsilon})$ edges, is NP-complete.
They also showed it is NP-complete to decide whether 
a graph with $e$ edges has an induced subgraph on $k$ vertices that has 
$\frac{ek^2}{n^2}(1+O(v^{\varepsilon-1}))$ edges; the latter
is only slightly larger than 
$\frac{ek^2}{n^2}(1-\frac{v-k}{vk-k})$, which is the 
the average number of edges in a subgraph with $k$ vertices.

For the case where edge weights fulfill the triangle inequality,
Ravi, Rosen\-krantz, and Tayi \cite{rrt-hscadp-94}
give a heuristic with time complexity
$O(n^2)$ and prove that it guarantees a performance bound of
4. (See Tamir \cite{t-cophscadp-98} with reference to this paper.)
Hassin, Rubinstein, and Tamir \cite{hrt-aamd-98}
give a different heuristic with time complexity $O(n^2+k^2\log k)$
with performance bound 2. On a related note, see
Chandra and Halld\'orsson \cite{chandra01approximation}, who study a
number of different remoteness measures for the subset $k$,
including total edge weight $w(S)$.
If the graph
from which a subset of size $k$ is to be selected is a tree, Tamir
\cite {t-oflg-91}
shows that an optimal weight subset can be determined
in $O(nk)$ time.

\newpage
In many important cases there is even more known about
the set of edge weights than just the validity of
triangle inequality. This is the case
when the vertex set $V$ corresponds to
a point set $P$ in geometric space, and distances between
vertices are induced by geometric distances between points.
Given the practical motivation for considering the problem,
it is quite natural to consider geometric instances of this type.
In fact, it was shown by Ravi, Rosenkrantz, and Tayi in
\cite{rrt-hscadp-94} that for the case of Euclidean distances
in two-dimensional space, it is possible to achieve 
performance bounds that are arbitrarily close to $\pi/2\approx 1.57$
For other metrics, however, the best performance guarantee
is the factor 2 by \cite{hrt-aamd-98}.
Despite of these approximation results,
it should be noted that the complexity status of the problem is still
open, i.e., it is known known whether the problem is NP-hard.

An important application of our problem is data sampling and clustering,
where points are to be selected from a large more-dimensional set.
Different metric dimensions of a data point
describe different metric properties of a corresponding item.
Since these properties are not geometrically related,
distances are typically not evaluated by Euclidean distances.
Instead, some weighted $L_1$ metric is used. (See Erkut \cite{erk-90}.)
For data sampling, a set of points is to be selected that has high
average distance.
For clustering, a given set of points is to be subdivided
into $k$ clusters, such that points from the same cluster are close
together, while points from different clusters are far apart.
If we do the clustering by choosing $k$ center points, and assigning
any point to its nearest cluster center, we have to consider
the same problem of finding a set
of center points with large average distance, which is equivalent
to finding a $k$-clique with maximum total edge weight.

For results on maximizing the {\em minimum} $L_1$
distance within a selected set of $n$ points 
see Baur and Fekete ~\cite{bf-agdp-01}, who showed that finding such a
set within a given polygon cannot be approximated arbitrarily well,
unless P=NP.
Finally, Gritzmann, Klee, and Larmann~\cite{gkl-ljsnp-95}
have studied a somewhat related geometric selection problem:
Given a set $v$ of $m$ points in $n$-dimensional space, choose
a subset of $n+1$ points, such that the total {\em volume}
of the resulting simplex is maximum. They showed that this problem
is NP-hard when $n$ is part of the input. 

\bigskip
\subsection*{Main Results}
In this paper, we consider point sets $P$ in $d$-dimensional
space, where $d$ is some constant. For the most part, distances are 
measured according to the rectilinear ``Manhattan'' norm $L_1$.

Our results include the following:

\begin{itemize}

\item
A linear time $(O(n))$ algorithm to solve the problem to optimality
in case where $k$ is some fixed constant. This is in contrast
to the case of Euclidean distances, where
there is a well-known lower bound of $\Omega(n\log n)$ 
in the computation tree model for
determining the diameter of a planar point set, i.e., the special
case $d=2$ and $k=2$ (see \cite{ps-cgi-85}).

\item
A polynomial time approximation scheme for the case where
$k$ is not fixed. This method can be applied for arbitrary
fixed dimension $d$. For the case of Euclidean distances
in two-dimensional space, it implies a performance bound
of $\sqrt{2}+\varepsilon$, for any given $\varepsilon>0$.
\end{itemize}

\section{Preliminaries}

For the most part of this paper,
all points are assumed to be points in the plane.
Spaces of arbitrary fixed dimension will be discussed in the end.
Distances are measured using the $L_1$ norm, unless noted otherwise.
The $x$- and $y$-coordinates of a point $p$ are denoted
by $x_p$ and $y_p$.
If $p$ and $q$ are two points in the plane,
then the distance between $p$ and $q$ is
$d(p,q) =  |x_p - x_q| + |y_p - y_q|$. We say that $q$ is above $p$
in direction of a vector $c$,
if the inner products satisfy
$\langle q,c\rangle\geq \langle p,c\rangle$
We say that a point $p$
is maximal in direction $c$ with respect to a set of points $P$
if it maximizes the inner product $\{\langle c,x\rangle\mid x\in P\}$. 
For example, 
if $p$ is an element of a set of points $P$ and $p$ has a maximal
$y$-coordinate, then $p$ is maximal in direction (0,1) with 
respect to $P$, and a point $p$ with minimal $x$-coordinate
is maximal in direction (-1,0) with 
respect to $P$. If the set $P$ is clear from the context, we simply
state that $p$ is maximal in direction $c$.

The weight of a set of points $P$ is the sum of the distances
between all pairs of points in this set, and is denoted
by $w(P)$. Similarly, $w(P,Q)$ denotes the total sum of distances
between two sets $P$ and $Q$. For $L_1$ distances, 
$w_x(P)$ and $w_x(P,Q)$ denote the sum of $x$-distances within $P$,
or between $P$ and $Q$.

\section{Cliques of Fixed Size}
\label{sec:fixed}

Let $S = \{s_0, s_1, \ldots, s_{k-1}\}$ 
be a maximum weight
subset of $P$, where $k$ is a fixed integer 
greater than 1. We will label the $x$- and $y$-coordinates
of a point $s \in S$ by some $(x_a,y_b)$ with $0 \leq a < k$
and $0 \leq b < k$ such that
$x_0 \leq x_1 \leq \ldots \leq x_{k-1}$ and
$y_0 \leq y_1 \leq \ldots \leq y_{k-1}$.
(Note that in general, $a\neq b$ for a point $s=(x_a,y_b)$.)
Then
\[  w(S) ~=~  \sum_{0 \leq i < j < k} (x_j - x_i) +
	       \sum_{0 \leq i < j < k} (y_j - y_i).   \]
		
Now we can use local optimality to reduce
the family of subsets that we need to consider:
%
%
%
%
%
\begin{lemma}
\label{le:sizek}
There is a maximum weight subset $S'$ of $P$ of cardinality $k$,
such that each point
in $S'$ is maximal in direction $(2i+1-k,2j+1-k)$ with respect to
$P\setminus S'$ for some values of $i$ and $j$ with $0 \leq i,j < k$.
\end{lemma}

\proof
\old{
Consider a maximum weight subset $S\subset P$ of cardinality
$k$. Let $s_i=(x_i,y_j)$ be a point in $S$ such that
there are $k-i-1$ points $s_\ell=(x_\ell,y_\ell)\in S\setminus\{s_i\}$
with $x_\ell>x_i$ (i.e., ``strictly to the right'' of $p$),
and $i$ points $s_\ell=(x_\ell,y_\ell)\in S\setminus\{s_i\}$
with $x_\ell\leq x_i$ (i.e., ``to the left'' of $p$).
Similarly, let there be $k-j-1$ points in $S\setminus\{s_i\}$
``strictly above`` $s_i$, and $j$ points in $S\setminus\{s_i\}$
``below'' $s_i$. Now consider replacing $s_i$ by a point
$s_i'=s_i+(h_x,h_y)$
with $(h_x,h_y)\geq 0$, i.e., above and to the right of $s_i$.
This increases the $x$-distance to the $i$ points
left of $s_i$ by $h_x$, while decreasing the $x$-distance to
the $k-i-1$ points right of $s_i$ by not more than $h_x$.
In the balance, replacing $s_i$ by $s_i'$ increases the total
$x$-distance by at least $(2i+1-k)h_x$.
Similarly, we get an increase of at least $(2j+1-k)h_y$
for the $y$-coordinates. If $s_i'$ is above $s_i$ in direction
$(2i+1-k,2j+1-k)$, the overall change is nonnegative,
and the claim follows.

Now consider the case in which there are several points
that are maximal in a given direction.
More precisely, suppose there is an $s_i'$ such that $(h_x,h_y) = (0,0)$.
Let $S' = S \setminus \{s\} \cup \{s_i'\}$.
Suppose point $s_i' = (x_i',y_i')$ is such that $x_i'$ and $y_i'$ have the same
rank in $S'$ as $x_i$ and $y_i$ have in $S$, i.e., there are $i$ points 
$s_l = (x_l,y_l) \in S' \setminus \{s_i'\}$ with $x_l \leq x_i'$
and  $j$ points $s_l = (x_l,y_l) \in S' \setminus \{s_i'\}$ with $y_l 
\leq y_i'$.
In this case $w(S') = w(S)$. However if $x_i'$ and $y_i'$ do not have the same
rank in $S'$ as $x_i$ and $y_i$ have in $S$, then it follows
that  $w(S') > w(S)$.
This contradicts the assumption that $w(S)$ is maximal. Therefore suppose that $S$ is a maximal
weight subset such that each point $s$ in $S$ is maximal in some direction 
$(2i+1-k,2j+1-k)$
with respect to $P - S$. If we replace $s$ by a point in $P-S$ that is maximal 
in the same direction, the weight of the subset remains the same.
}
Consider a maximum weight subset $S \subset P$ of cardinality $k$.
Let $s = (x_i,y_{i'})$  be a point in $S$, such that
there are $i$ points $s_l = (x_l,y_{l'}) \in S \setminus \{ s \}$
with $x_l \leq x_i$ (i.e., to the left of $s$)
and $k-i-1$ points $s_l = (x_l,y_{l'}) \in S \setminus \{ s \}$
with $x_l > x_i$ (i.e., strictly to the right of $s$).
Similarly let there be $j$ points below $s$ and $k-j-1$ points strictly
above $s$. We claim that $s$ is maximal in direction $(2i+1-k,2j+1-k)$
with respect to $P-S$.

Consider replacing $s$ by a point $s' = s + (x_h,y_h)$ in $P-S$.
Let $ \delta = (2i+1-k)x_h + (2j+1-k)y_h$. 
Let $S' = S \setminus \{s\} \cup \{s'\}$.
Assume first that point $s' = (x_i',y_{i'}')$ is such that 
$x_i'$ and $y_{i'}'$ have the same
rank in $S'$ as $x_i$ and $y_{i'}$ have in $S$, i.e., there are $i$ points
$s_l = (x_l,y_l) \in S' \setminus \{s'\}$ with $x_l \leq x_i'$
and  $j$ points $s_l = (x_l,y_{l'}) \in S' \setminus \{s'\}$ with $y_{l'} \leq y_{i'}'$.
Replacing $s$ by $s'$ changes the $x$-distances to the points left of $s$
by $ix_h$, and the $x$-distances to the points right of $s$
by $(k-i-1)(-x_h)$. Similarly, the $y$-distances change by
$jy_h$ and $(k-j-1)y_h$.
So $w(S') = w(S) + (2i+1-k)x_h + (2j+1-k)y_h = w(S) + \delta$.
Since $w(S)$ is maximum, we derive that $\delta \leq 0$,
i.e.,  no point in $P - S$ is above any point in $S$ in 
direction $(2i+1-k,2j+1-k)$.

If the $x$- and $y$-coordinates of $s'$ do not have the same rank in $S'$
as the $x$- and $y$-coordinates of $s$ in $S$, then it is not hard to show that
$w(S') > w(S) + \delta$, so $\delta < 0$. Therefore in this case, $s'$ is 
strictly below $s$ in direction $(2i+1-k,2j+1-k)$. 

We can also conclude that if $s = (x_i,y_{i'})$ and $s' = (x_i',y_{i'}')$
are at the same level in direction $(2i+1-k,2j+1-k)$, i.e., if $\delta = 0$,
then the $x$- and $y$-coordinates of $s'$ do have the same rank in $S'$
as the $x$- and $y$-coordinates of $s$ in $S$ and $w(S) = w(S')$.
\qed

\begin{theorem}
Given a constant value for $k$, a maximum weight subset $S$ of 
a set of $n$ points $P$, such that $S$ has 
cardinality $k$, can be found in linear time.
\end{theorem}

\proof
Consider all directions of the form $(2i+1-k,2j+1-k)$ 
with $0 \leq i,j < k$.
For each direction $(a,b)$, find  
$S_k (a,b)$, a set of $k$ points that  are maximal in direction $(a,b)$
with respect to $P - S_k (a,b)$. 
Compute the set $\cup S_k (a,b)$ and try
all possible subsets of size $k$ of this set
until a subset of maximum weight is found.

Correctness follows from the fact that Lemma \ref{le:sizek} implies
that $S \subset \cup S_k (a,b)$. Since $k$ is a constant, each set
$S_k (a,b)$ can be found in linear time. Since the cardinality 
of $\cup S_k (a,b)$ is less than or equal to $k^3$, the result follows.
{From} the discussion at the end of the proof
of Lemma~\ref{le:sizek} we can conclude that if the set of $k$ points
 maximal
in a direction $(a,b)$ is not unique, any set of $k$ points maximal
in this direction will work equally well.

\qed

\bigskip
Note that in the above estimate, we did not try
to squeeze the constants in the $O(n)$ running time.
A closer look shows that for $k=2$, not more than
$2$ subsets of $P$ need to be evaluated for possible optimality,
for $k=3$, 8 subsets are sufficient.

\newpage
\section{Cliques of Variable Size}

In this section we consider the scenario where
$k$ is not fixed, i.e., $k$ is part of the input.
We show that there is a polynomial time approximation scheme
(PTAS), i.e., for any fixed positive $\eps$, there is a
polynomial approximation algorithm that finds a solution that
is within $(1+\eps)$ of the optimum.

\begin{figure}[htb]
 \begin{center}
  \leavevmode
  \centerline{\epsfig{file=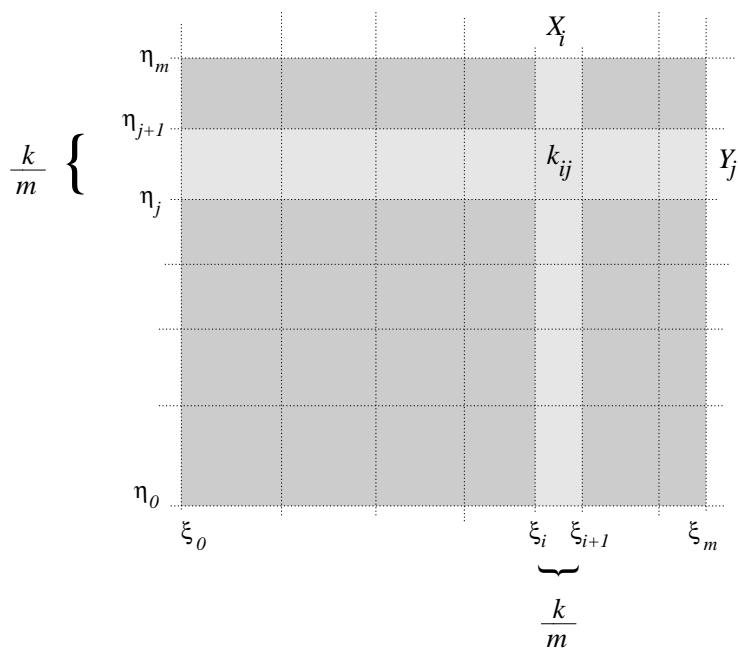, 
  width=0.8\textwidth}}
  \caption{Subdividing the plane into cells}
  \label{fig:cells}
 \end{center}
\end{figure}

The basic idea is to use for each of the $d$ coordinates
a suitable subset of $m_\varepsilon$
coordinate values that subdivide an optimal solution into subsets
of equal cardinality. More precisely, we describe the case $d=2$;
we find (by enumeration) a subdivision of an optimal solution into
$m_\varepsilon\times m_\varepsilon$ rectangular cells $C_{ij}$, 
each of which must
contain a specific number $k_{ij}$ of selected points.
{From} each cell $C_{ij}$, the points are selected in a way that
guarantees that the total distance to all other cells
except for the $m_\varepsilon-1$ cells in the same ``horizontal'' strip
or the $m_\varepsilon-1$ cells in the same ``vertical'' strip
is maximized. As it turns out, this can
be done in a way that the total neglected distance within the 
strips is bounded by a fraction of $(5m_\varepsilon-9)/(2(m_\varepsilon-1)(m_\varepsilon-2))$ of the 
weight of an optimal solution, 
yielding the desired approximation
property. See Figure~\ref{fig:cells} for the overall picture.

For ease of presentation we assume that $k$ is a multiple of $m_\varepsilon$ and
$m_\varepsilon > 2$.
Approximation algorithms for other values of $k$ can be constructed
in a similar fashion. 
Consider an optimal solution
of $k$ points, denoted by $\OPT$. 
Furthermore consider a division of the plane by
a set of $m_\varepsilon+1$ $x$-coordinates 
$\xi_{0}\leq\ldots\leq\xi_{1}\leq \xi_{m_\varepsilon}$. 
Let $X_i:=\{p=(x,y)\in\Re^2 \mid \xi_{i}\leq x\leq \xi_{i+1}, 0 \leq i < m_\varepsilon \} $
be the vertical strip between coordinates $\xi_i$ and $\xi_{i+1}$.
By enumeration of possible choices of $\xi_{0},\ldots,\xi_{m_\varepsilon}$
we may assume that the $\xi_i$
have the property that, for an optimal solution, from each of the 
$m_\varepsilon$ strips $X_i$ precisely $k/m_\varepsilon$ points of $P$
are chosen.
(A small perturbation does not change optimality
or approximation properties of solutions. This shows that in case
of several points sharing the same coordinates, ties
may be broken arbitrarily; in that case, points on
the boundary between two strips may be considered belonging
to one or the other of those strips, whatever is convenient
to reach the appropriate number of points in a strip.)

In a similar manner, suppose we know $m_\varepsilon+1$ $y$-coordinates
$\eta_0 \leq \eta_1 \leq \ldots \leq \eta_{m_\varepsilon}$ 
such that
from each horizontal strip 
$Y_i:=\{p=(x,y)\in\Re^2 \mid \eta_i\leq y\leq \eta_{i+1}, 0 \leq i < m_\varepsilon  \}$
a subset of $k/m_\varepsilon$ points are chosen for
an optimal solution.

Let $C_{ij}:=X_i\cap Y_j$, and let $k_{ij}$
be the number of points in $\OPT$ that are chosen from $C_{ij}$.
Since $\sum_{0 \leq i < m_\varepsilon} k_{ij}=
\sum_{0 \leq j < m_\varepsilon} k_{ij}=k/m_\varepsilon$, we may assume by enumeration
over the $O(k^m_\varepsilon)$
possible partitions of $k/m_\varepsilon$ into $m_\varepsilon$ pieces
that we know all the numbers $k_{ij}$.

Finally, define the vector
$\nabla_{ij}:= ((2i+1-m_\varepsilon)k/m_\varepsilon,(2j+1-m_\varepsilon)k/m_\varepsilon)$.
Now our approximation algorithm is as follows: from 
each cell $C_{ij}$, choose some $k_{ij}$ points that
are maximal in direction $\nabla_{ij}$.
(Overlap between the selections from different cells is avoided 
by proceeding in lexicographic order of cells, and choosing the 
$k_{ij}$ points among the candidates that are still unselected.)
Let \HEU\ be the point set selected in this way.

It is clear that $\HEU$ can be computed in polynomial time.
We will proceed
by a series of lemmas to determine how well $w(\HEU)$ approximates
$w(\OPT)$. 
In the following, we consider the distances involving points
from a particular cell $C_{ij}$. 
Let $\HEU_{ij}$ be the set of 
$k_{ij}$ points that are selected from $C_{ij}$ by the heuristic,
and let $\OPT_{ij}$ be a set of $k_{ij}$ points 
of an optimal solution that are attributed to $C_{ij}$. 
Let $S_{ij}=\OPT_{ij}\cap \HEU_{ij}$.
Furthermore we define $\ol{S}_{ij}=\HEU_{ij}\setminus\OPT_{ij}$,
and $\tilde{S}_{ij}=\OPT_{ij}\setminus\HEU_{ij}$.
Let $\HEUip$, $\OPTip$, $\HEUpj$ and $\OPTpj$ be the set of $k/m_\varepsilon$ points selected
from $X_i$ and $Y_j$ by the heuristic and an optimal algorithm respectively.
Finally
$\ol{\HEU}_{i\bullet}:=\HEU\setminus\HEUip$,
$\ol{\HEU}_{\bullet j}:=\HEU\setminus\HEUpj$,
$\ol{\OPT}_{i\bullet}:=\OPT\setminus\OPTip$
and $\ol{\OPT}_{\bullet j}:=\OPT\setminus\OPTpj$.

\begin{lemma}
\label{le:external}
\begin{eqnarray*}
   w_x(\HEUij,\notHEUip) + w_y(\HEUij,\notHEUpj)\\
   \geq w_x(\OPTij,\notOPTip) + w_y(\OPTij,\notOPTpj).  
\end{eqnarray*}
\end{lemma}

\proof
Consider a point $p \in \tilde{S}_{ij}$. Thus, there is
a point $p'\in\ol{S}_{ij}$ that was chosen by the heuristic 
instead of $p$. Now we can argue
like in Lemma~\ref{le:sizek}: Let $h=(h_x,h_y)=p'-p$.
When replacing $p$ in $\OPT$ by $p'$, we 
increase the $x$-distance to the $ik/m_\varepsilon$ points
``left'' of $C_{ij}$
by $h_x$, while decreasing the $x$-distance to
$(m_\varepsilon-i-1)k/m_\varepsilon$ points ``right'' of $C_{ij}$ by $h_x$. 
In the balance, this yields a change of $((2i+1-m_\varepsilon)k/m_\varepsilon)h_x$.
Similarly, we get a change of $((2j+1-m_\varepsilon)k/m_\varepsilon)h_y$
for the $y$-coordinates. By definition, we have assumed that 
the inner product $\langle h,\nabla_{ij}\rangle\geq 0$, so the overall
change of distances is nonnegative.

Performing these replacements for all points in
$\OPT\setminus\HEU$, we can transform $\OPT$ to $\HEU$, while
increasing the sum of distances 
$w_x(\OPTij,\notOPTip) + w_y(\OPTij,\notOPTpj)$ 
to the sum 
$w_x(\HEUij,\notHEUip) + w_y(\HEUij,\notHEUpj)$.

\qed

\bigskip
In the following three lemmas we show that the total difference between
the weight of an optimal solution $w(\OPT)$ and the total value 
of all the right hand sides (when summed over $i$) of
the inequality in Lemma \ref{le:external} is a small fraction
of $w(\OPT)$.

\begin{lemma}
\label{le:middlestrips}
\begin{eqnarray*}
   \sum_{0<i<m_\varepsilon-1} w_x(\OPTip) ~\leq~ \frac{w_x(\OPT)}{2(m_\varepsilon-2)}. 
\end{eqnarray*}
\end{lemma}

\proof
Let $\delta_i = \xi_{i+1} - \xi_i$.
Since $i(m_\varepsilon-i-1)\geq m_\varepsilon-2$ for $0<i<m_\varepsilon-1$,
we have for $0 < i < m_\varepsilon-1$
\[ w_x(\OPTip) \leq \frac{k^2}{2m_\varepsilon^2} \delta_i 
               \leq \frac{ik}{m_\varepsilon} \frac{(m_\varepsilon-i-1)k}{m_\varepsilon} \delta_i ~\frac{1}{2(m_\varepsilon-2)}. \]
Since $\OPT$ has $ik/m_\varepsilon$ and $(m_\varepsilon-i-1)k/m_\varepsilon$ points to the left
of $\xi_i$ and right of $\xi_{i+1}$ respectively, we have
\[ w_x(\OPT) \geq \sum_{0<i<m_\varepsilon-1} \frac{ik}{m_\varepsilon} \frac{(m_\varepsilon-i-1)k}{m_\varepsilon} \delta_i \]
so
\[ \sum_{0<i<m_\varepsilon-1} w_x(\OPTip)  \leq \frac{1}{2(m_\varepsilon-2)} w_x(\OPT). \]
\qed

\begin{lemma}
\label{le:endstrips}
For $i = 0$ and $i=m_\varepsilon-1$ we have
\[ w_x(\OPTip) ~\leq~ \frac{w_x(\OPT)}{m_\varepsilon-1} \]
\end{lemma}

\proof
Without loss of generality assume $i=0$.
Let $x_0, x_1, \cdots , x_{(k/m_\varepsilon)-1}$ be the $x$-coordinates of the 
points $p_0,p_1,\ldots,p_{(k/m_\varepsilon)-1}$ in $\OPTop$. So 
\[ w_x(\OPTop) ~=~  (\frac{k}{m_\varepsilon}-1)(x_{\frac{k}{m_\varepsilon}-1}-x_0) 
                  + (\frac{k}{m_\varepsilon}-3)(x_{\frac{k}{m_\varepsilon}-2}-x_1) 
               +~ \ldots \]
\[ \leq (\frac{k}{m_\varepsilon}-1)(\xi_1-x_0) + (\frac{k}{m_\varepsilon}-3)(\xi_1-x_1) +~ \ldots \]
\[ \leq \frac{k}{m_\varepsilon} (\xi_1-x_0) + \frac{k}{m_\varepsilon}(\xi_1-x_1) +~ \ldots \] 
Since $\xi_1 - x_j \leq x - x_j$ where $0 \leq j < k/m_\varepsilon$ and
$x$ is the $x$-coordinate of any point in $\notOPTop$
and since there are $(m_\varepsilon-1)k/m_\varepsilon$ points
in $\notOPTop$, we have
\[ \xi_1 - x_j < \frac{m_\varepsilon}{(m_\varepsilon-1)k} w_x (p_j,\notOPTop) \]
so
\begin{eqnarray*}
w_x(\OPTop)  &\leq& \frac{k}{m_\varepsilon} \frac{m_\varepsilon}{(m_\varepsilon-1)k}
\sum_{0 \leq i < \frac{k}{2m_\varepsilon}} w_x (p_i,\notOPTop) \\
&\leq& \frac{1}{m_\varepsilon-1} \sum_{0 \leq i < \frac{k}{m_\varepsilon}} w_x (p_i,\notOPTop) \\
&=&    \frac{1}{m_\varepsilon-1} w_x (\OPTop, \notOPTop) \\
&\leq& \frac{1}{m_\varepsilon-1} w_x (\OPT) .
\end{eqnarray*}
\qed

This proves the main properties. Now we only have to combine the
above estimates to get an overall performance bound:

\begin{lemma}
\label{le:allstrips}
\begin{eqnarray*}
   \sum_{0 \leq i < m_\varepsilon} w_x(\OPTip,\notOPTip) + 
   \sum_{0 \leq j < m_\varepsilon} w_y(\OPTpj,\notOPTpj) \\
   \geq (1 - \frac{5m_\varepsilon-9}{2(m_\varepsilon-1)(m_\varepsilon-2)}) w(\OPT)  ).
\end{eqnarray*}
\end{lemma}

\proof
{From} Lemmas \ref{le:middlestrips} (applied for indices $0<i<m-1$)
and \ref{le:endstrips} (applied twice, once for $i=0$, and once for $i=m-1$),
we derive that
\[ \sum_{0\leq i < m_\varepsilon} w_x(\OPTip) ~\leq~ \frac{5m_\varepsilon-9}{2(m_\varepsilon-1)(m_\varepsilon-2)} w_x(\OPT) \]
and similarly
\[ \sum_{0\leq i < m_\varepsilon} w_y(\OPTpj) ~\leq~ \frac{5m_\varepsilon-9}{2(m_\varepsilon-1)(m_\varepsilon-2)} w_y(\OPT). \]
Since
\begin{eqnarray*}
w(\OPT) &=& w_x(\OPT) + w_y(\OPT)  \\
&=&  \ \ \ \sum_{0 \leq i < m_\varepsilon} w_x(\OPTip,\notOPTip) +
\sum_{0 \leq i < m_\varepsilon} w_x(\OPTip)    \\
&&+
\sum_{0 \leq j < m_\varepsilon} w_y(\OPTpj,\notOPTpj) +
\sum_{0 \leq j < m_\varepsilon} w_y(\OPTpj),
\end{eqnarray*}

the result follows.
\qed

\bigskip
Putting together  Lemma \ref{le:external} and the error estimate from Lemma
\ref{le:allstrips}, the approximation
theorem can now be proven.

\begin{theorem}
\label{th:ptas}
For any fixed $m$, \HEU\ can be computed in polynomial time,
and 
\[ w(\HEU) \geq (1-\frac{5m_\varepsilon-9}{2(m_\varepsilon-1)(m_\varepsilon-2)}) w(\OPT).  \]
The running time is exponential in $\frac{1}{\varepsilon}$.
\end{theorem}

\proof
The claim about the running time is clear. 
(The only step that is exponential in $\frac{1}{\varepsilon}$
is the enumeration over all $O(k^{m_\varepsilon})$
possible partitions of $k/m_\varepsilon$ into $m_\varepsilon$ pieces.)
Using Lemmas \ref{le:external}
and \ref{le:allstrips}
we derive
\[ w(\HEU) \geq \sum_{0 \leq i <  m_\varepsilon} w_x(\HEUip,\notHEUip)
              + \sum_{0 \leq j <  m_\varepsilon} w_y(\HEUpj,\notHEUpj) \]
\[ \geq \sum_{0 \leq i <  m_\varepsilon} w_x(\OPTip,\notOPTip)
      + \sum_{0 \leq j <  m_\varepsilon} w_y(\OPTpj,\notOPTpj) \]
\[ \geq (1 - \frac{5m_\varepsilon-9}{2(m_\varepsilon-1)(m_\varepsilon-2)}) w(\OPT).  \]
\qed

\section{Implications}
It is straightforward to modify our above arguments
to point sets under $L_1$ distances
in an arbitrary $d$-dimensional space, with fixed $d$.

\begin{theorem}
Given a constant value for $k$ and $d$, a maximum weight subset $S$ of 
a set of $n$ points in $d$-dimensional space, such that $S$ has 
cardinality $k$, can be found in linear time.
If $d$ and $\eps$ are constants, 
but $k$ is not fixed,
then there is a polynomial time algorithm
that finds a subset whose weight
is within $(1+\eps)$ of the optimum.
\end{theorem}

For the case of fixed $k$, it is straightforward to generalize the argument
from Section~\ref{sec:fixed} to see that there are at most $(2k)^d$
interesting directions to consider. For $k$ being part of the input,
the approximation scheme can be generalized in a straightforward manner
by using an $m_\varepsilon$-subdivision in each coordinate direction.
Again the complexity ends up being exponential in $\frac{1}{\varepsilon}$,
as well as in $d$.

For the case of $L_\infty$ distances in the plane, the results for
$L_1$ distances can be applied by a standard argument: A rotation
by $\pi/4$ transforms $L_\infty$ distances into $L_1$ distances and vice versa.
Furthermore, we can use the approximation scheme from
the previous section to get a $\sqrt{2}(1+\varepsilon)$ approximation
factor for the case of Euclidean distances in two-dimensional space,
for any $\varepsilon>0$: In polynomial time,
find a $k$-set $S_1$ such that $L_1(S)$
is within $(1+\varepsilon)$ of an optimal solution $\OPT_1$
with respect to $L_1$ distances. Let $\OPT_2$ be an optimal
solution with respect to $L_2$ distances. Then
\begin{eqnarray*}
L_2(S)&\geq& \frac{1}{\sqrt{2}} L_1(S)
\geq \frac{1}{\sqrt{2}(1+\varepsilon)} L_1(\OPT_1)
\geq \frac{1}{\sqrt{2}(1+\varepsilon)} L_1(\OPT_2)\\
&\geq& \frac{1}{\sqrt{2}(1+\varepsilon)} L_2(\OPT_2),
\end{eqnarray*}
and the claim follows. Similarly, any norm has its characteristic
approximation factor $\rho$ with respect to $L_1$ or $L_\infty$ distances; 
this factor immediately yields a $(\rho+\varepsilon)$-approximation
for geometric dispersion.

\section{Conclusions}
We have presented algorithms for geometric instances of
the maximum weighted $k$-clique problem. Our results
give a dramatic improvement over the previous best
approximation factor of 2 that was presented
in \cite{hrt-aamd-98} for the case of general metric spaces.
This underlines the observation
that geometry can help to get better algorithms for 
problems from combinatorial optimization. 

Furthermore, the algorithms in
\cite{hrt-aamd-98} give better performance for Euclidean
metric than for Manhattan distances. 
We correct this anomaly by showing
that among problems involving geometric distances,
the rectilinear metric may allow better algorithms than
the Euclidean metric. 

It remains an interesting open problem to show NP-hardness
of a geometric version of the problem for spaces of fixed dimension.
In particular, the case of Manhattan distances in the plane
may actually turn out to be polynomially solvable.

\section*{Acknowledgments}
We would like to thank Katja Wolf and Magn\'us Halld\'orsson
for helpful discussions, Rafi Hassin for several
relevant references, and three anonymous referees for useful
comments.

\newpage
\bibliographystyle{plain}
\bibliography{refs}  

\end{document}